\documentclass[a4paper,11pt]{article}
\pdfoutput=1 

\usepackage{jcappub} 

\usepackage[T1]{fontenc} 
\usepackage{graphicx}
\usepackage{bm}
\usepackage[utf8]{inputenc}
\usepackage{amssymb}
\usepackage{tensor}
\usepackage{xspace} 
\usepackage{tabularx}
\usepackage{multirow}
\usepackage{dsfont}
\title{\boldmath Non-locality in  Quadrupolar Gravitational Radiation}

\author[a,b,c]{S. Capozziello,}
\author[d]{M. Capriolo,}
\author[e]{A. Carleo}
\author[d,f]{G. Lambiase}

\affiliation[a]{Dipartimento di Fisica "E. Pancini",  Universit\`{a}  degli Studi di Napoli Federico II", Via Cinthia, I-80126, Napoli, Italy}
\affiliation[b]{Istituto Nazionale di Fisica Nucleare (INFN), sez. di Napoli, Via Cinthia 9, I-80126 Napoli, Italy}
\affiliation[c]{Scuola Superiore Meridionale, Largo S. Marcellino, I-80138, Napoli, Italy}
\affiliation[d]{Dipartimento di Fisica "E.R. Caianiello",   Universit\`{a}  di Salerno, Via Giovanni Paolo II, I-84084 Fisciano (SA), Italy.}
\affiliation[e]{INAF, Osservatorio Astronomico di Cagliari, I-09047, Selargius (CA), Italy}
\affiliation[f]{INFN - Gruppo Collegato di Salerno, Italy}

\emailAdd{capozziello@na.infn.it}
\emailAdd{mcapriolo@unisa.it}
\emailAdd{amodio.carleo@inaf.it}
\emailAdd{lambiase@sa.infn.it}

\abstract{General Relativity suffers for two main problems which have not yet been overcome: it predicts spacetime singularities and cannot be formulated as a perturbative renormalizable theory. In particular, many attempts have been made for avoiding singularities, such as considering higher order  or infinite derivative theories. The price to pay in both cases is to give up locality and therefore they  are known altogether as non-local  theories of gravity. In this paper, we investigate how to recognize the presence of non-local effects by exploiting the power emitted by gravitational waves in a binary system in presence of  non-local corrections as  $R\Box^{-1}R$ to the Hilbert-Einstein action. After solving the field equations in terms of the source stress-energy tensor $T_{\mu\nu}$ and obtaining the gravitational wave stress-energy pseudo-tensor,   $\tau_{\mu\nu}$, we find that the General Relativity quadrupole formula is modified in a non-trivial way, making it feasible to find a possible signature of non-locality.  Our final results on the    gravitational wave stress-energy pseudo-tensor   could also be applied to several  astrophysical scenarios involving energy or momentum loss, potentially providing multiple tests for non-local deviations from General Relativity. We finally  discuss the detectability of the massless transverse scalar mode, discovering that, although  this radiation is  extremely weak, in a small range around the model divergence, its amplitude could fall within the low-frequency Einstein Telescope sensitivity.   }

\begin{document}
\maketitle
\flushbottom


\section{Introduction}
\label{sec1}
\setcounter{equation}{0}
Modern physics is based on two main pillars, i.e.
General Relativity and Quantum Field Theory. General Relativity (GR) describes the gravitating systems and non-inertial frames
on large scales, while  Quantum Field
Theory (QFT) holds at high energy or, equivalently,  small scale regimes where a classical description
breaks down. However, QFT assumes that space-time is
flat and even its extensions, such as QFT in curved space
time, consider space-time as a classical background, never allowing a quantized version of the latter. GR, on the other hand, does not take into account the quantum nature of matter because the source of the gravitational field simply is the
(classical) Stress-Energy tensor $T_{\mu\nu}$ and hence, in its formulation, gravity is a \textit{local} interaction. Although a theory of Quantum Gravity remains unknown, it is legitimate to ask for what happens when a strong gravitational field is present at quantum scales. Because of its weakness compared to  other interactions,  the characteristic scale under which one would
expect to experience full non-classical effects, relevant to gravity, the Planck
scale is quite small, i.e. $10^{-33}$ cm, which is  not  accessible by any current
experiment. \\

Furthermore, the Einstein Equivalence  Principle   is based on the
assumption that an accelerated observer in Minkowski spacetime, at each
event along its world line, is physically equivalent to a momentarily identical
comoving inertial observer. Non-locality  comes out taking also
the past history of the accelerated observer into account. Such history-dependent theories  have recently
been developed with the purpose of see a quantum feature arising from GR, while addressing,  at the same time, some of the most important issues of modern Cosmology, namely dark energy or dark matter.  It is not a new fact that Quantum Mechanics shows non-local aspects: non-locality is a manifestation of entanglement and the latter  has been repeatedly demonstrated in laboratory experiments; the Bell theorem, furthermore, demonstrated that locality is violated in some quantum systems. After that, non-locality  has been investigated in  QFT (see e.g. \cite{Pauli:1949,Pauli:1953,Efimov:1,Efimov:2,Efimov:3,Efimov:4,Efimov:5}), as well as in string theory  \cite{String_1,String_2,String_3,string_4}, where it mainly  emerges as a 'side' effect of the introduction of quantum corrections to GR the purpose of which was to cure the singularity problem. Indeed, applications to Cosmology showed that non-local ghost-free higher-derivative
modifications of the Einstein gravity, in the ultraviolet regime, can  admit non-singular bouncing solutions for the Universe (in place of the Big Bang singular  solution) and non-singular Schwarzschild metrics for black holes \cite{Modesto_2011}. Another motivation for considering non-local gravity is the possibility of achieving renormalizability without the appearance of ghost modes \cite{Modesto_Renormalizab}.  
A road towards
Quantum Gravity, by  considering non-local corrections to the Hilbert-Einstein action, has been drawn in  \cite{Modesto:2015lna,Modesto:2015ozb,Nojiri:2019dio,Nojiri:2007uq}. The introduction of non-local terms have also  been considered in alternative theories of gravity, such as teleparallel gravity \cite{Bahamonde:2017bps} and revealed extremely useful in cosmology \cite{Capozziello:2024qol, Nojiri:2010pw, Capozziello:2008gu}.  However, it must be said that  the non-local theories of gravity  themselves can be  considered as Extended Theories of Gravity \cite{Capozziello:2011et}, meaning  that when the non-local terms are cancelled out,   the GR limit is recovered.\\
More precisely, non-local theories of gravity are described by Lagrangians composed by a finite sum of products between fields and
their derivatives evaluated at different points $x$ and $y$
of the spacetime while metric $g_{\mu\nu}$ and/or other fields  are
described by integro-differential equations, implying that the value of the field at one point depends on its
value at another point of the spacetime, weighted by a function called nucleus or kernel. There are essentially three different ways to implement non-locality from a mathematical point of view. The first approach (the most studied in the literature) is by means of a convergent series expansion with real coefficients $c_n$ of an analytic non-polynomial
function $F$ of the d'Alembert operator $\Box$, known as Infinite Derivative  Gravity  (IDG)\cite{Efimov:1,Buoninfante_1,Buoninfante_2,Tesi_IDG_Edholm}:
\begin{equation*}
    F(\Box) = \sum_{n=1}^{\infty} c_n \dfrac{\Box^n}{M^{2n}}\ ,
\end{equation*}
where $M$ is the mass scale associated to non-locality effects. Notice that, in this case,  non-locality appears only at the level of interaction, and not at the level of a free field propagation without a source. Recently, it was shown that, in gravity theories containing such   class of non-local terms,
the linearized Ricci tensor $R_{\mu\nu}$ and  Ricci scalar $R$ are not vanishing in the region of non-locality, i.e. at short distance from a source, due to the smearing of the source induced by the presence
of non-local gravitational interactions. It follows that, unlike in the Einstein gravity, the
Riemann tensor is not traceless and it does not coincide with the Weyl tensor, which, however, vanishes
at short distances, implying that the (static) metric is  conformally flat in that region
\cite{Buoninfante_1}, implying a possible deviation from $1/r$ potential drop at very short distance \footnote{A decay as $1/r$ of the gravitational potential has only been verified up to $\sim 10^{-5}$ m, which is thirty orders of magnitude away from the Planck length \cite{bilancia_tors,bilancia_2}.}. 

In the second way, that we shall follow here, the non-locality  manifests itself in non-analytic operators such as \cite{Capriolo,complex}
\begin{equation}\label{0_1}
    \Box^{-n} R(x) = \phi_{0}(x) + \int_{\Omega} d^4y \sqrt{-g(y)} G(x,y) R(y)\ ,
\end{equation}
where $G(x,y)$ is the retarded Green function of the operator $\Box^{n}$, $\Omega \subseteq \mathbb{R}^{4}$ and $\phi_0$ is the solution of $\Box^{-n}\phi_0 (x)=0$.  It was shown that the application of the non-local operator $\Box^{-1}$
to the
scalar curvature $R$ gives rise to the late-time cosmic expansion of the Universe without invoking any Dark Energy contribution. For an overview on non-local cosmology see also \cite{Review_Non_local, Bouche2022, Capozziello:2024qol}. Finally, non-locality can enter  through a constitutive relation on the (linearized) gravitational field involving a causal kernel  determined via observational data, in the spirit of non-local electrodynamics of media \cite{Mashhoon:2022,Puetzfeld_2019}.

In all these approaches, it is important to study the linearized versions of the theories and to
derive gravitational waves (GWs). For IDG, they have been studied in \cite{GWs_IDG_2012} and \cite{GWs-IDG_2018}, while, for higher order theories with Lagrangian  $\mathcal{L}= R + \sum_{h=1}^{n} a_h R \Box^{-h} R$, with $n$ fixed, they are discussed in \cite{Capriolo}. Indeed, gravitational radiation allows to detect  possible effects
of non-local gravity \cite{Capcap:2020xem} as well as to classify the degrees of freedom of a given theory.   \\

In this paper, we  compute the gravitational wave stress-energy tensor  (GW-SET) for the non-local gravity theory $\mathcal{L}= R + a R \Box^{-1} R$ with $a$ an adimensional constant and derive, from it,  the power emitted by a gravitational system in terms of  quadrupole momenta, in view of possible astrophysical applications. This gravitational action is a peculiar case of models considered in  \cite{DW_original}, whose cosmological implications have been  studied in \cite{Amendola_2019}. A modified quadrupole formula with non-local terms has been studied for binary pulsars, see~\cite{CARLEO2024138410}. A derivation of gravitational pseudo-tensor in higher order curvature-based and  torsion-based gravity has been taken into account in~\cite{https://doi.org/10.1002/andp.201600376,Capozziello:2018qcp}. In particular, in Sec. \ref{sec2} we obtain the field equations and the Noether current using a localized version of the Lagrangian, in order to avoid integro-differential equations. Then, in Sec. \ref{sec3},  we solve the non-homogeneous system of equations in terms of the (source) stress-energy tensor $T_{\mu\nu}$, whose solutions in the far region condition only depends on the quadrupole momenta. Finally, we close the manuscript with a summary and an outlook in Sec. \ref{sec4}. More detailed accounts of some computations are given in the appendices. \\
In this work, we adopt  natural units  $\hbar=c=1$, and we define the Planck
mass  as $M^{2}_{pl} = {8 \pi G}$. The negative metric signature  $(+, -, -, -)$ is also adopted. The Greek indices are the coordinate ones contracted with the metric tensor $g_{\mu \nu}$, while the Latin indices (spatial coordinates) are   contracted with the Kronecker delta. Finally, space vectors are indicated in bold.


\section{Non-local gravity as an extension of General Relativity}
\label{sec2}
Let us derive the field equations for an extended theory of gravity given by the action 
\begin{equation}\label{action}
 S[g] = \dfrac{1}{2 \chi} \int d^4 x \sqrt{-g} R\Big( 1+ f(\Box^{-1} R)  \Big) + \int d^4 x \sqrt{-g} \mathcal{L}_m[g]\ ,
\end{equation}
where  $\chi= 8 \pi G$ is the gravitational coupling, $\mathcal{L}_m$ is the matter Lagrangian,  and $f(\Box^{-1})$ is an analytic function of $\Box^{-1}$ which is the inverse of  d'Alembert operator $\Box = \nabla^{\mu}\nabla_{\mu}$, being $\Box^{-1} \Box = \mathds{1}$. A similar action has been considered in  \cite{DW_original}. According to the form of $f(\Box^{-1})$, it shows  Noether symmetries   \cite{S2}.  An interesting case presents  a non-local correction like $\sim R  \exp \{a \Box^{-1}R\}$ where $a$ is an adimensional constant. However, we can simplify the previous action by expanding $f$ in terms of $\Box^{-1}R$  and considering only the first term, which, in the quadrupole approximation  we are going to discuss, is the dominant one. It is
\begin{equation}
f(\Box^{-1})\approx a\Box^{-1}R\ .
\end{equation}
Therefore the action~\eqref{action} becomes \cite{2107.06972} 
\begin{equation}\label{action4}
 S[g] = \dfrac{1}{2 \chi} \int d^4 x \sqrt{-g} \Big( R + aR \Box^{-1} R  \Big) + \int d^4 x \sqrt{-g} \mathcal{L}_m[g]\ ,
\end{equation}
In view to derive the field equations, we can recast  action  (\ref{action}) in a {\it localized} version. It is

\begin{equation}\label{action2}
S_g [g_{\mu\nu},\phi,\lambda] = \dfrac{1}{2 \chi} \int d^4 x \sqrt{-g} \Big[ R (1+a\phi  ) + \lambda(\Box\phi - R) \Big]\ ,
\end{equation}

where the auxiliary field $\phi = \Box^{-1} R$ has been introduced. Here   $\lambda$ is a Lagrange multiplier. See Ref.\cite{Review_Non_local} and references therein for details.
After integration by parts, Eq. (\ref{action2}) becomes 

\begin{equation}
S [g_{\mu\nu},\phi,\lambda] = \dfrac{1}{2 \chi} \int d^4 x \sqrt{-g} \Big[ R (1+a\phi - \lambda ) - \nabla^{\alpha} \lambda \nabla_{\alpha}\phi \Big].
\end{equation}

The variation with respect to the metric $g^{\mu\nu}$  gives the  equations of motion, i.e.
\begin{equation}\label{eq:fe}
    G_{\mu\nu} + (G_{\mu\nu} + g_{\mu\nu} \Box - \nabla_{\mu} \nabla_{\nu}  ) (a\phi - \lambda ) - \nabla_{(\mu}\phi \nabla_{\nu)}\lambda + \dfrac{1}{2} g_{\mu\nu} \nabla^{\sigma}\phi\nabla_{\sigma}\lambda = \chi T_{\mu\nu}\ ,
\end{equation}
while variations with respect to both the scalar fields give the two constraints 
\begin{equation}\label{constr}
    \Box \lambda = -a R \;, \; \; \; \; \Box \phi =  R\ .
\end{equation}
The stress-energy tensor is $T_{\mu\nu} = - \dfrac{2}{\sqrt{-g}}\dfrac{ \delta (\sqrt{-g}\mathcal{L}_m)}{\delta g^{\mu\nu}}$, and  the Einstein tensor is $G_{\mu\nu}= R_{\mu\nu}-\dfrac{1}{2}g_{\mu\nu}R$.
Tracing Eq. (\ref{eq:fe}), one obtains the scalar equation  
\begin{equation}
    (1+a\phi-6a-\lambda) R - \nabla^{\sigma}\phi\nabla_{\sigma}\lambda  = - \chi T \ ,
\end{equation}
where $T=g^{\mu\nu}T_{\mu\nu}$ and we used the relation 
\begin{equation}\label{rel}
    \Box \big( a \phi - \lambda  \big) = 2 a R \ .
\end{equation}
An important issue is whether it is possible to eliminate or not the non-local terms from the Lagrangian~\eqref{action4} by redefining the metric tensor as done, for example,  in Ref.~\cite{deRham:2020ejn}.  This is not possible for  the following mathematical and physical reasons.  In fact, from a mathematical point of view, non-locality is linked to the integro-differential character of the field equations and therefore, from Eq. ~\eqref{0_1}, the non-local operators are integral operators where the value of a physical quantity at a point depends on its value assumed at other points modulated by the kernel.  Thus it is not possible to eliminate the integral character of physical quantities through a redefinition of the metric which is a local tensor object. On the other hand, from a physical point of view, non-locality is in this theory an intrinsic property of gravity related to  finite characteristic lengths and masses. Also if a "localization" process is performed (see e.g. \cite{Nojiri:2007uq, Review_Non_local} )  auxiliary scalar fields emerge pointing out that non-locality introduces further degrees of freedom.

In order to analyze the gravitational radiation, we  consider the first-order perturbations $h_{\mu\nu}$, $\varphi$ and $w$. The metric perturbation is   around the flat metric $\eta_{\mu\nu}$ and  the perturbations of the two scalar fields  $\phi$ and $\lambda$ are around their (constant) Minkowskian values $\phi_0$ and $\lambda_0$, respectively, i.e. 
\begin{equation}
\begin{array}{cc}
     g_{\mu\nu} = \eta_{\mu\nu} + h_{\mu\nu} \;  , \\
     \phi = \phi_0 +  \varphi \; , \\
     \lambda = \lambda_0 + w  \; .
\end{array}
\end{equation}

At first order in $h_{\mu\nu}$ and without imposing any gauge, the Ricci tensor $R_{\mu\nu}$ and Ricci scalar $R$ are  \cite{Carroll:2004st}
\begin{equation}
    R^{(1)}_{\mu\nu}=\dfrac{1}{2}\Big(\partial_{\rho}\partial_{\mu}h^{\rho}_{\nu} + \partial_{\rho}\partial_{\nu}h^{\rho}_{\mu} -\partial_{\mu}\partial_{\nu} h - \Box h_{\mu\nu} \Big) \;  ,
\end{equation}

\begin{equation}\label{R1}
    R^{(1)} = \partial_{\mu}\partial_{\nu} h^{\mu\nu} -\Box h \; .
\end{equation}
In GR, the gravitational pseudo-tensor can be obtained in different ways,  however,  the  related physical information has to be the same. This statement holds also in modified gravity  as showed in \cite{princeton}. Here, we  follow the Noether current method, based on the  symmetries of the action, which lead to
conserved currents. The gravitational stress-energy tensor  comes out  by varying, on shell, the second-order Lagrangian density $\mathcal{L}^{(2)}$  with respect to the coordinates. In our case, it is 
 
\begin{equation}\label{eq:sec}
    \mathcal{L}^{(2)} = \dfrac{1}{2 \chi} \Big[ (\sqrt{-g}R)^{(2)} \Big(1+a \phi_{0} -\lambda_{0}\Big) + R^{(1)} \Big(a\varphi -w\Big) - \partial^{\mu}\varphi \partial_{\mu}w \Big] \; .
\end{equation}

Defining $\psi_{0} \doteq 1 +a \phi_{0} -\lambda_{0}\neq 0$ and $\psi^{(1)} \doteq a \varphi - w$ its perturbation, and introducing the new field 

\begin{equation}\label{gauge}
    \theta_{\mu\nu} \doteq  h_{\mu\nu} - \dfrac{1}{2} \eta_{\mu\nu} h - \dfrac{\eta_{\mu\nu}}{\psi_{0}}\psi^{(1)} ,
\end{equation}
such that, in our reference frame, it is $\partial_{\mu}\theta^{\mu\nu}=0$ (the Lorenz gauge),  Eq. (\ref{eq:sec}) becomes

\begin{equation}\label{L2_final}
\begin{array}{cc}
    \mathcal{L}^{(2)} = \dfrac{\psi_{0}}{32 \pi G } \Big[ \dfrac{1}{4} \partial_{\alpha} \theta \partial^{\alpha} \theta + \partial_{\alpha}\theta_{\beta \gamma }  \partial^{\beta}\theta^{\alpha \gamma }  - \dfrac{1}{2} \partial_{\gamma} \theta_{\alpha\beta} \partial^{\gamma} \theta^{\alpha\beta} + \dfrac{(6a-2\psi_{0})}{\psi_{0}^{2}} \partial^{\alpha}\varphi \partial_{\alpha} w   \\
    - \dfrac{3}{\psi_{0}^2} \Big( a^2\partial_{\alpha}\varphi \partial^{\alpha}\varphi + \partial_{\alpha}w \partial^{\alpha}w   \Big)   \Big]\ . 
    \end{array}
\end{equation}
See Appendix A for details. 
From this Lagrangian, the Euler-Lagrange equations of motion, in the Lorenz gauge and in vacuum,  are

\begin{equation}\label{eq:eomtheta}
\Box \theta_{\mu\nu} = 0\ ,
\end{equation}
\begin{equation}\label{eq:eomphi}
K \Box w - \dfrac{6 a}{\psi_0^2} \Box \varphi = 0\ , 
\end{equation}
\begin{equation}\label{eq:eomw}
K \Box \varphi - \dfrac{6 a}{\psi_0^2} \Box w = 0\ ,
\end{equation}
as already found in \cite{2107.06972}. In Eqs. (\ref{eq:eomphi}) and (\ref{eq:eomw}), we have defined the constant $K \doteq \dfrac{(6a-2\psi_{0})}{\psi_{0}^{2}}$. Notice that the last  two are coupled differential equations for $\Box \varphi$ and $\Box w$. When $a=0$ they implies $\Box \varphi = 0 = \Box w$ and therefore $\varphi= 0 = w$, that is  GR  is recovered.    The trivial solution of the system formed by the equations \eqref{eq:eomphi} and \eqref{eq:eomw} is obtained by imposing that the determinant is different from zero, that is 
\begin{equation}\label{eqfieldscalar1}
6a\neq\psi_{0}, 
\end{equation}
which implies 
\begin{equation}\label{eqfieldscalar2}
\Box \varphi=\Box w=0 , 
\end{equation}
or
\begin{equation}\label{eqfieldscalar3}
 \Box \psi^{(1)}=0\,. 
\end{equation}
This means that   $\phi^{(1)}$ is a massless scalar field  added to the two tensor fields of the field Eq.~\eqref{eq:eomtheta}.  In other words,  gravity linearly perturbed presents  three degrees of freedom. 

The   Noether current for our  non-local gravity model is defined as 
\begin{equation}\label{curr}
    j^{\alpha}_{\beta} \doteq - \dfrac{\partial \mathcal{L}^{(2)}}{\partial (\partial_{\alpha} \theta_{\mu\nu} )  } \partial_{\beta} \theta_{\mu\nu} - \dfrac{\partial \mathcal{L}^{(2)}}{\partial (\partial_{\alpha} \varphi)} \partial_{\beta} \varphi - \dfrac{\partial \mathcal{L}^{(2)}}{\partial (\partial_{\alpha} w)} \partial_{\beta} w + \delta_{\beta}^{\alpha} \mathcal{L}^{(2)}\ ,
\end{equation}
which, after some computations (see Appendix A), becomes

\[\arraycolsep=1.4pt\def\arraystretch{2.2}
\begin{array}{cc}
   j^{\alpha}_{\beta} = \dfrac{\psi_{0}}{32 \pi G} \Big[   -2 \partial^{\nu}\theta^{\mu\alpha} \partial_{\beta} \theta_{\mu\nu} + \partial^{\alpha}\theta^{\mu\nu}\partial_{\beta}\theta_{\mu\nu} - K\partial^{\alpha} w \partial_{\beta}\varphi 
   \\ 
   + \dfrac{6}{\psi_{0}^2} a^2 \partial^{\alpha}\varphi \partial_{\beta}\varphi - K\partial^{\alpha} \varphi \partial_{\beta}w 
   + \dfrac{6}{\psi_0^2} \partial^{\alpha} w \partial_{\beta}w \\
+\delta^{\alpha}_{\beta} \partial_{\nu}\theta_{\mu\gamma} \partial^{\gamma} \theta^{\mu\nu} - \dfrac{1}{2} \delta^{\alpha}_{\beta} \partial_{\gamma}\theta_{\mu\nu} \partial^{\gamma} \theta^{\mu\nu} + \delta^{\alpha}_{\beta} K \partial^{\gamma} \varphi \partial_{\gamma}w - \delta^{\alpha}_{\beta} \dfrac{3 a^2}{\psi_0^2} \partial^{\gamma} \varphi \partial_{\gamma}\varphi - \delta^{\alpha}_{\beta}  \dfrac{3 }{\psi_0^2} \partial^{\gamma} w \partial_{\gamma}w  \Big]\ , 
\end{array}
\]

where we have already dropped terms with $\theta$ since they would be zero because of the TT gauge. For similar reasons, the first and the  fifth-from-last terms are zero \footnote{For our purpose, the Noether current should be considered under average over several wavelenghts, $\langle j^{\alpha}_{\beta} \rangle $, in order to obtain a gauge-invariant measure of physical quantities which we are interested in (such as energy and momentum). This imply that single derivatives are zero, $\langle \partial_{\mu} f \rangle = 0$, and hence we are empowered to integrate by parts under the the averaging brackets. }, while the fourth  is zero thanks to the equation of motion  (\ref{eq:eomtheta}). Furthermore, the last three terms in the above equation are also zero. Integrating by parts, they can be recast as 
\begin{equation}\label{eq:comb}
\begin{array}{cc}
   \delta^{\alpha}_{\beta} \Big[  K \partial^{\gamma} \varphi \partial_{\gamma}w -  \dfrac{3 a^2}{\psi_0^2} \partial^{\gamma} \varphi \partial_{\gamma}\varphi -  \dfrac{3 }{\psi_0^2}\partial^{\gamma} w \partial_{\gamma}w  \Big] \\
   = \delta^{\alpha}_{\beta} \Big[  - K  \varphi \Box w  + \dfrac{3 a^2}{\psi_0^2}  \varphi \Box\varphi + \dfrac{3 }{\psi_0^2}  w \Box w  \Big] \\
   = \delta^{\alpha}_{\beta} \Big[  - \dfrac{1}{2} K  \varphi \Box w  - \dfrac{1}{2} K  w \Box \varphi   + \dfrac{3 a^2}{\psi_0^2}  \varphi \Box\varphi + \dfrac{3 }{\psi_0^2}  w \Box w  \Big]\ , 
   \end{array}
\end{equation}
where, in the last line, we have simply  written $K\varphi \Box w$ as the sum of two equal terms and integrated the second one. Then, the final result in Eq. (\ref{eq:comb}) can be seen as a combination\footnote{Multiplying  Eq. (\ref{eq:eomphi}) by a factor $-\varphi / 2 $ and Eq. (\ref{eq:eomw}) by a factor $-w / 2 $, the sum is equal to Eq. (\ref{eq:comb}). } of Eqs. (\ref{eq:eomphi}) and (\ref{eq:eomw}) and therefore it is always zero. By deleting all the null terms, what remains is

\begin{equation}\label{eq:2.19}
    j^{\alpha }_{\beta } = \dfrac{\psi_0}{32 \pi G } \Big[   \partial^{\alpha} \theta^{\mu\nu}  \partial_{\beta} \theta_{\mu\nu} + \dfrac{6}{\psi_0^2} \partial^{\alpha}\psi^{(1)} \partial_{\beta} \psi^{(1)} + 
    \dfrac{4}{\psi_0} \partial^{\alpha}\varphi \partial_{\beta} w \Big]\ ,
\end{equation}
which holds only in the TT gauge and under averaging brackets. \\
The gravitational wave stress-energy tensor (GW-SET) is then defined as $\tau^{\alpha}_{\beta} \doteq \langle j^{\alpha}_{\beta} \rangle$, therefore 

\begin{equation}\label{pseudo_final}
    \tau^{\alpha}_{\beta} = \dfrac{\psi_0}{32 \pi G } \Big\langle   \partial^{\alpha} \theta^{\mu\nu}_{(TT)}  \partial_{\beta} \theta_{\mu\nu}^{(TT)} + \dfrac{6}{\psi_0^2} \partial^{\alpha}\psi^{(1)} \partial_{\beta} \psi^{(1)} + 
    \dfrac{4}{\psi_0} \partial^{\alpha}\varphi \partial_{\beta} w \Big\rangle \ .
\end{equation}

where we have stressed the gauge condition on $\theta_{\mu\nu}$ and brackets means average over all wavelenghts. Notice that the above expression  is symmetric in the indices  $\alpha$, $\beta$ and it is not invariant under diffeomorphisms, but only under affine transformations. In this sense, it is a pseudo-tensor. Therefore, to make it a real  tensor, the average procedure is essential.  Furthermore, as we can see, in addition to  GR term (the first) and a scalar field contribution from $\psi^{(1)}=a\varphi-w$, a mixed term also appears.  It is not possible to recast it as a  term involving only $\psi$.   We would expect a pseudo-tensor including only terms in $\theta_{\mu\nu}$  and $\psi^{(1)}$ as in the case of scalar-tensor theories \cite{princeton} or $f(R)$ gravity \cite{hiroshima}. The non-zero mixed term is a peculiarity of non-local gravity which could constitute a signature for such theories.

 As a final remark, we point out that  combining  relations in  Eqs. (\ref{constr}),  it turns out that 
\begin{equation}
\Box (\lambda+a\psi)=0 , 
\end{equation}
and then the first-order perturbation gives 
 \begin{equation}
\Box (w+a\psi)=0 ,
\end{equation}
whose trivial solution is $w=-a\psi$.  This means that the two fields $w$ and $\psi$ can be chosen as redefinitions of each other and that, therefore, it is necessary to add only one additional degree of freedom to the two standard  tensor degrees of freedom of  linearized GR.  Based on this, the $f(R)$-like field $\psi^{(1)}$ can be written   as $\psi^{(1)}=2a\varphi$. It is clear, from the above considerations and in the following sections, that  the three scalar fields are related each other and non-local gravity implies only a further mode in GWs \cite{Capcap:2020xem}.

\section{Gravitational Waves with Source}
\label{sec3}

In order to get physical quantities from the pseudo-tensor (\ref{pseudo_final}), we have to link each of the fields $\theta_{\mu\nu}, \varphi, w$ to the source of gravitational waves, that is to $T_{\mu\nu}$. This implies to solve the equations of motions not in vacuum, but with $T_{\mu\nu}\not=0$. The field equations with source,  in the Lorentz gauge,  are \footnote{Since our background space is Minkowski, we have $\phi_{0}= \Box^{-1} R^{(0)} = 0 $ and, similarly, $\lambda_{0}=0$. Furthermore, notice that in solving non-vacuum equations,  the traceless condition is not allowed.   } 

\begin{equation}\label{eq:pieno1}
    \Box \theta_{\mu\nu} = -2 \chi  T_{\mu\nu}\ ,
\end{equation}
\begin{equation}\label{eq:pieno2}
\Box \varphi = \dfrac{1}{2} \Box \theta + 3 \Box \psi^{(1)} \ ,
\end{equation}
\begin{equation}\label{eq:pieno3}
\Box w = - a \left[ \dfrac{1}{2} \Box \theta + 3 \Box \psi^{(1)} \right]\ .
\end{equation}
Here the matter energy-momentum tensor $T_{\mu\nu}$ stands for the unperturbed one because the continuity equation $\partial_{\mu}T^{\mu\nu}=0$ must be fulfilled. The proportionality between $\Box \varphi$ and $\Box w$ is obvious and it comes directly from Eq. (\ref{constr}).
The trace equation $\Box \theta = - 2 \chi T$ can be recast in the form

\begin{equation}\label{eq:trace1}
    \Box \psi ^{(1)} = \dfrac{1}{3} \Big[ \psi_0 R^{(1)} + \chi T  \Big]\ ,
    \end{equation}
where we have used $G^{(1)} = - R^{(1)} $ with (see Eq. (\ref{R1}))
\begin{equation}\label{eq:R1}
    R^{(1)} = -\dfrac{1}{2}\Box h + \dfrac{1}{\psi_0}\Box \psi^{(1)}\ .
\end{equation}
with $\psi_0 = 1$. Eq. (\ref{eq:trace1}) is analogous to the trace equation of $f(R)$ gravity  where it  is usually recasted as a Klein-Gordon equation for the scalar field \cite{Capozziello:2011et}. We will see  later that this is not the case for  non-local gravity, i.e. it is not possible to obtain a Klein-Gordon equation for the scalar field $\psi^{(1)}$. Notice also that  Eqs. (\ref{eq:pieno2}) and (\ref{eq:pieno3}) are not identical to Eqs. (\ref{eq:eomphi}) and (\ref{eq:eomw}) since now $\Box \theta \not=0$.\\
The above  non-homogeneous linear system of equations has two possible solutions according to the value of $a$. For $a\neq 1/6$, the linearized field equation admits the standard Einsten gravitational waves plus massless transverse scalar radiation while the non-local $\Box^{-1}$-theory, for the value of the coupling constant $a=1/6$, shows a different behavior. It admits massive scalar waves as solutions and reproduces the Polyakov two-dimensional conformal effective action that occurs in string theory. See  \cite{complex, Capriolo}. If $a=1/6$, it is:
\begin{equation}\label{sol_1}
  a=1/6 \; \; \; \; \; \; \; \Rightarrow \; \; \; \; \; \; \; \Bigl\{  T=0 \; ,  \; \; \Box w = - \dfrac{1}{6} \Box \varphi  \Bigr\}\ .
\end{equation}
The traceless condition at zero order $T_{\mu\nu}$ is not suitable for our purposes: it is not physically acceptable for a binary system where the linear approximation holds  and, more importantly, it would not allow to  obtain a solution for the fields $\varphi$ and $w$ depending on  the source. In other words, a contribution from the non-local term $a R \Box^{-1}R$ to the quadrupole formula of the gravitational radiation directly generated by the matter is not admissible if $a=1/6$ for the slowly moving compact binaries, where  the background space-time curvature is negligible. However, this fact   does not exclude the presence of lower multipoles (monopole and dipole) in gravitational radiation in addition to the quadrupole and higher multipole contributions, in presence of  strong-field sources. Hence, we will not consider this case any more. \\

On the other hand, when  $a\not=1/6$,  the solution of the system is the following: 

\begin{equation}\label{sol_2}
a\not=1/6 \; \; \; \; \; \; \; \Rightarrow \; \; \; \; \; \; \;  \Bigl\{ \Box \varphi = \dfrac{1}{6a-1} \chi T \; , \; \Box w = \dfrac{-a}{6a-1} \chi T   \Bigr\}\ ,
\end{equation}
which, once solved in turn,  clearly give explicit solutions for the scalar fields $\varphi$, $w$ in terms of the source. Differently from the case $f(R)$, (\ref{sol_2}) are not standard Klein-Gordon equations, because no mass term appears, according to the fact that, for $a\neq 1/6$,  the non-local $\Box^{-1}$-gravity  shows only massless scalar modes but not massive ones. One might see the effect of non-locality in the manifestation of an effective energy-momentum tensor, namely $\Tilde{T}\doteq - T/[2(6a-1)].$  Moreover, since $\Box \psi^{(1)} = 2 a \Box \varphi $, getting solutions from (\ref{sol_2}) is equivalent to find a solution for $\psi^{(1)}$ as well. Based on these facts,  in the following we will focus only on the case (\ref{sol_2}).

\subsection{Explicit Solutions}
We want now to seek for explicit solutions for $\theta_{\mu\nu}$, $\varphi$ and $w$ with the assumption of far wave zone, i.e. for a source far enough from the observer, and we also assume a source in non-relativistic motion in order to simplify the orbital  description of  the binary system. \\
In analogy with GR, Eq. (\ref{eq:pieno1})  immediately returns
\begin{equation}\label{eq:38}
    \theta_{\mu\nu}(t,\mathbf{x}) = 4 G \int \dfrac{T_{\mu\nu}(t-|\mathbf{x}-\mathbf{y}|,\mathbf{y})}{|\mathbf{x}-\mathbf{y}|} d^3\mathbf{y} := J_{\mu\nu}\ ,
\end{equation}

which, in the hypothesis of far-field, i.e. $|\mathbf{x}-\mathbf{y}|^{-1} \simeq 1/r$, becomes 

\begin{equation}
 \theta_{ij}(t,\mathbf{x}) \simeq \dfrac{4 G}{r} \int T_{ij}(t-r,\mathbf{y}) d^3\mathbf{y}\ . 
\end{equation}
In the above equation, we considered only spatial components since the remaining ones can be computed exploiting the gauge condition as in GR. We call $t' \doteq t - r$ the retarded time and, using the conservation of energy-momentum $\partial_{\nu}T^{i \nu}=0$ and the  relation
\begin{equation}
\partial_{\mu}\partial_{\nu} \left( x^{i}x^{j}T^{\mu\nu} \right) = 2 T^{i j} ,
\end{equation}
 Eq. (\ref{eq:38}) turns into
\begin{equation}\label{theta_quadr}
   \theta_{ij}(t,\mathbf{x}) = \dfrac{2G}{r} \dfrac{d^2 Q_{ij}(t')}{d^2 t}\ ,
\end{equation}

where we have defined the quadrupole tensor 
\begin{equation}\label{quadrupoleQ}
    Q_{ij}(t) \doteq \int d^3 y T^{00}(t,\mathbf{y}) y^i y^j\ ,
\end{equation}
with integration over the entire space of the source \footnote{In linearized gravity, $T_{\mu\nu}$ is a zero-order quantity. This implies that the contribution to the energy-momentum tensor comes only from ordinary matter and not from  gravitational waves. It turns out that $T_{\mu\nu}$ is zero outside and on the boundary of the source. }. The above quantity  is closely linked to the geometry and masses involved in the astrophysical system and it  is therefore a known quantity. Solution (\ref{theta_quadr}), on the other hand, gives the gravitational tensor field as a function of the quadrupole time-derivative, to be replaced  into Eq. (\ref{pseudo_final}). \\
From Eq. (\ref{sol_2}), it turns out that  the relation $w=-a\varphi$ holds, and therefore it is sufficient to obtain a solution for only one scalar field. More precisely they are

\begin{equation}\label{phi_integrale}
    \varphi(t) \simeq \dfrac{2G}{(1-6a)r}\int T(t-r,\mathbf{y})d^3\mathbf{y}=\dfrac{\eta^{\mu\nu}}{2(1-6a)}J_{\mu\nu} \; ,
\end{equation}

\begin{equation}\label{w_integral}
w(t) \simeq  - \dfrac{2aG}{(1-6a)r}\int T(t-r,\mathbf{y})d^3\mathbf{y}=- \dfrac{a \eta^{\mu\nu}}{2(1-6a)}J_{\mu\nu} \; .  
\end{equation}
In obtaining the above solutions, we assumed the solution of the homogeneous equations to be zero, as in GR. In other words, we are assuming that when the source is off, all the fields are zero, just as for $h_{\mu\nu}$, and this is legitimate if we want such fields to be generated by matter.  Apart from this,  when $a\not= 1/6$,  the GW-SET (\ref{pseudo_final}) reduces to 

\begin{equation}\label{pseudo_final_new}
    \tau^{\alpha}_{\beta} = \dfrac{\psi_0}{32 \pi G } \Big\langle   \partial^{\alpha} \theta^{\mu\nu}_{(TT)}  \partial_{\beta} \theta_{\mu\nu}^{(TT)} + 
    4a(6a-1) \partial^{\alpha}\varphi \partial_{\beta} \varphi \Big\rangle\ , 
\end{equation}

where only one scalar field appears. Notice that the above expression is always  symmetric and gauge-invariant \footnote{The first term, corresponding to the tensor contribution, is invariant under gauge transformation only considering the average procedure, as in GR. The second term, i.e. the non-local correction, is manifestly covariant. Explicitly, under a transformation ${x'}^{\mu} = x^{\mu} + \epsilon^{\mu}(x) $, it turns out that $\partial_{\mu}\varphi \rightarrow \partial_{\mu}\varphi + \partial_{\sigma}\varphi \partial_{\mu} \epsilon^{\sigma}$, and therefore   $\partial_{\alpha}\varphi \partial_{\beta}\varphi \; \;   \rightarrow \; \;  \partial_{\alpha}\varphi \partial_{\beta}\varphi + \partial_{\alpha}\varphi \partial_{\rho}\varphi \partial_{\beta} \epsilon^{\rho} + \partial_{\beta}\varphi \partial_{\rho}\varphi \partial_{\alpha} \epsilon^{\rho}$, i.e. $\partial_{\alpha}\varphi \partial_{\beta}\varphi$ transforms like a 2-rank tensor.}.   The trace $J$ in Eq. (\ref{phi_integrale}) can be developed as  
\begin{equation}\label{J}
    J(t) = \dfrac{4G}{r} \Big[M+n^i \Dot{D}_i(t')+\dfrac{1}{2}n_in_j\Ddot{Q}^{ij}(t')\Big] + \dfrac{2G}{r}\Ddot{Q}(t')\ ,
\end{equation}
with $Q \doteq \delta_{ij}Q^{ij} $, overdot means derivative with respect to time $t'$, and $n_i\doteq \hat{x}_i = x_i/r$.  See also \cite{Laurentis_2011}. Therefore, using conservation laws $\Dot{M}$ and $\Ddot{D}^i=0$, we arrive at (see Appendix B)
\begin{equation}\label{part_part}
    \partial_0 \varphi \partial_r \varphi = \dfrac{-1}{4(1-6a)^2}\left[ \dfrac{2G}{r}\left( n_i n_j \dddot{Q}^{ij}(t')+\dddot{Q} \right) \right]^2 \; .
\end{equation}

\subsection{The Emitted Power}
The quadrupole radiation is  the emitted power (or luminosity) in the far region condition, and it is given by  \cite{Maggiore:I}
\begin{equation}\label{P_tot}
    P_{tot} = - r^2 \int d\Omega \langle \tau^{0 i} \rangle  n_i := P_{tot}^{(GR)} + P_{tot}^{(NL)}\ ,
\end{equation}
with integration on a spatial surface at spatial infinity. The subscript 'NL' labels the non-local contribution in addition to the GR result, which corresponds to the first term of Eq. (\ref{pseudo_final_new}). It is well-known that the latter is given by \cite{Carroll:2004st,Maggiore:I} 
\begin{equation}\label{powerGR}
    P_{tot}^{(GR)} = \dfrac{G}{5 c^5} \langle \dddot{Q}_{ij} \dddot{Q}^{ij} - \dfrac{1}{3} \dddot{Q}^2 \rangle  \; ,
\end{equation}

where we  have restored the factor $c^5$ at the denominator. Using Eq. (\ref{part_part}), after long but straightforward calculations, the non-local contribution to the energy rate reads as 

\begin{equation}\label{power_final_NL}
    P_{tot}^{(NL)} = \dfrac{a G}{15 (6a-1)} \langle  \dddot{Q}^{jk}\dddot{Q}_{jk} + 13 \dddot{Q}^2 \rangle\ ,
\end{equation}
which  clearly cancels out when $a=0$. Substituting Eqs. (\ref{powerGR}) and (\ref{power_final_NL}) into  Eq. (\ref{P_tot}), the following  formula for quadrupole radiation in non-local gravity is found: 
\begin{equation}\label{p_tot_finale}
  \boxed{  P_{tot} = \dfrac{G}{5 c^5} \Big[ \Big(1+ \dfrac{a}{3 (6a-1)} \Big) \langle \dddot{Q}^{ij}\dddot{Q}_{ij} \rangle + \dfrac{1+7a}{3(6a-1)} \langle \dddot{Q}^2 \rangle   \Big]  } \; .
\end{equation}
Just to give a preliminary example, 
 we can consider  a Keplerian binary system consisting in two stars of mass $M$ in a circular orbit at distance $R$ from their common center of mass, then Eq. (\ref{p_tot_finale}) predicts an emitted power of 
\begin{equation}\label{power_case_1}
    P_{tot} = \dfrac{128 G M^2 R^4 w^6}{5 c^5} \left[ 1 - \dfrac{a}{3(1-6a)} \right]\ ,
\end{equation}
where $w=\sqrt{\frac{GM}{4R^3}}$ is the keplerian angular frequency of the binary system. In general, for binary systems with different masses  $m_{1}$ and $m_{2}$, on elliptic orbits of semi-major axis $R$ and eccentricity $e$,  from the  Peters - Mathews formula  \cite{Maggiore:I}, it turns out that  

\begin{equation}
    P_{tot} = \dfrac{32 G^4 m^2 \mu^2}{5 c^5 R^5} f(e,a)\ ,
\end{equation}
where  $ m= m_{1}+m_{2}$ is the total mass and $ \mu=m_{1}m_{2}/(m_{1}+m_{2})$ is the reduced mass, and we have defined the function 

\begin{equation}
    f(e,a) \doteq \dfrac{e^4(241 a-37)+4 e^2 (469 a -73) + 608 a-96}{96(6a-1)(1-e^2)^{7/2}}\ .
\end{equation}

From Eq. (\ref{power_case_1}) it is evident that non-local gravity induces a modification in quadrupole approximation of emitted power from binary systems, therefore resulting eventually observable. Since we expect a  small value for $a$ (any correction to GR in any extended theory of gravity should be rather slight), the  above result  shows that non-local corrections are, at least in principle, compatible with orbit decays by quadrupole  radiation in binary systems, paving the way for new observational constraints on non-local corrections compared to those already present in the literature \cite{Amendola_2019, Bouche2022, S2}. Of course, to find constraints on the non-local parameter $a$ we need to apply Eq. (\ref{p_tot_finale})  to realistic astrophysical systems. This way to proceed, i.e. constraining parameters after the selection of  the functional form for the correction, is an alternative to a more general approach, where the functional form itself can be selected by fitting  the observations. \\

\subsection{Amplitude of the scalar mode}

Finally, let us evaluate the amplitude of the gravitational wave. As mentioned before, the tensor mode of the gravitational wave is given by the field $\theta_{i j}$ defined in Eq. (\ref{gauge}). From an observational point of view, it would be desirable  to separate the purely metric contribution $h_{ij}$ from the field contribution given by $\psi^{(1)}= a \varphi -  w$, since we just have Eq. (\ref{theta_quadr}) which holds for the full field $\theta_{i j}$. For this purpose, one could assume $\varphi$ and $w$ as distinct propagating (scalar) modes, according to Eqs. (\ref{phi_integrale}) and (\ref{w_integral}),  where we have

\begin{equation}\label{JJ}
    \eta^{\mu\nu} J_{\mu\nu} \equiv  J \simeq \dfrac{4 G}{r} \Big[ M +  \dfrac{1}{2} \Ddot{Q} \Big]\ ,
\end{equation}
approximating Eq. (\ref{J}) considering only ($1/r$)-terms. Therefore, $\varphi$ and $w$ could be seen as breathing
modes in this theory of gravity. In order to estimate the amplitude of such modes, we first notice that the trace of the quadrupole tensor for a binary system is given by $Q=L^2 \mu$, where we called $L=2 R$ the separation between the two bodies. Here $\mu$ is the reduced mass. Based on this assumption, a scalar contribution will be present only if $L$ is a function of time, as assumed in \cite{hiroshima}. In particular, considering a trend like $L=L_0 \big(1-\frac{t}{t_{coal}}\big)^{1/4}$, it turns out that 

\begin{equation}
    \Ddot{Q}(t) = - \dfrac{1}{4}\dfrac{\mu L_0^2}{t_{coal}^2} \Big( 1- \dfrac{t'}{t_{coal}}\Big)^{-3/2} \ ,
\end{equation}
where $t_{coal}$ is the coalescence time and we have evaluated the quantity at  the retarded time $t'=t-r/c$. Halfway through the life of the binary system, i.e. at $t'=t_{coal}/2$, the modulus will be \footnote{The monopole term, proportional to the (total) mass $M$, in Eq. (\ref{JJ}) is static and therefore it is negligible w.r.t. the quadrupole radiation. } 

\begin{equation}
    |\varphi| \sim \dfrac{G \mu L_0^2 }{r (1-6a)c^4  t_{coal}^2}\ ,
\end{equation}
which clearly depends on the non-local parameter $a$. For estimating the strain, we first need an estimate of the coalescence time in non-local gravity. Assuming a circular orbit\footnote{The following argument  could easily be generalized to elliptical orbits.  We also denote with $M_{tot}$ the total mass of the binary system in order to generalize to the case of different masses.} for simplicity, the period is $T=2\pi \sqrt{\frac{R^3}{G M_{tot}}}$ then

\begin{equation}
    \dfrac{\Dot{T}}{T} = \dfrac{3}{2}\dfrac{\Dot{R}}{R} \;\; \; \; \rightarrow \; \; \; \;   \Dot{R}(t) = - \Tilde{k}R^{-3}(t)\ ,
\end{equation}
where we define 
\begin{equation}
\Tilde{k}\doteq \frac{192}{15}\frac{G^3 \mu M_{tot}^2}{c^5}f(e,a)\ .
\end{equation}

\begin{figure}
	\centering 
	\includegraphics[width=1.0\textwidth]{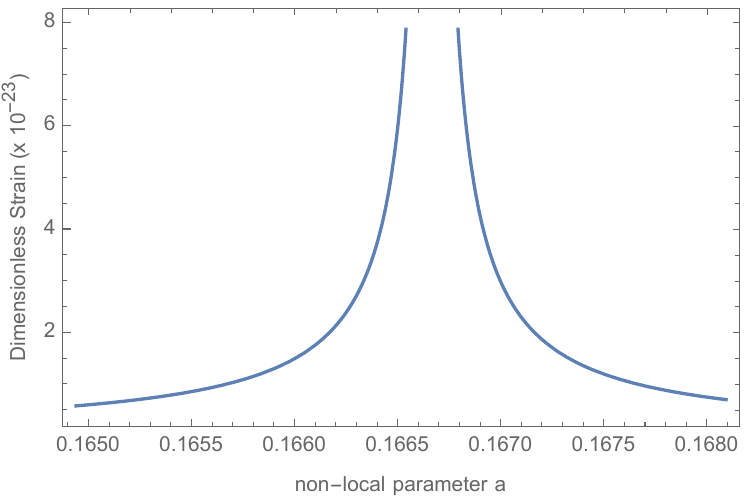}	
	\caption{Dimensionless strain of the scalar mode for a black-hole binary system with masses $100 M_{\odot}$  and orbital distance of $0.01$ AU as a function of the dimensionless non-local parameter $a$. In this range of values, the non-local massless transverse scalar mode could fall within the low frequency  sensitivity of the Einstein Telescope. } 
	\label{fig1}
\end{figure}

In obtaining the above equation for $R(t)$, we used the relation between the variation of the period and the variation of the gravitational energy, i.e.  $\frac{\Dot{T}}{T} = - \frac{3}{2}\frac{\Dot{E}}{E}$, where $E=-(G M \mu) /(2 R)$ and $\Dot{E}=-P_{tot}$ given in the previous section.  For circular orbits,  $f(e,a) \approx 1$ in analogy with the GR case, hence the solution of the above differential relation is 
\begin{equation}
R(t)=\sqrt{2}(c_1-\Tilde{k}t)^{1/4}\ , 
\end{equation}

where $c_1$ is found by implying that $R(0)=R_{now}$, with $R_{now}$ the current value for $R$. It turns out that $c_1=R^4_{now}/4$. With this value, the coalescence time, $t_{coal}$, is the time at which  $a(t_{coal})=0$, giving 

\begin{equation}
   t_{coal}= \dfrac{5}{256}\dfrac{c^5 R_{now}^4}{G^3 \mu M_{tot}^2 }\ . 
\end{equation} 

which is  the well-known result in GR. This is enough to obtain an order-of-magnitude estimate for $t_{coal}$ valid also for the present non-local theory. In particular, for two black holes with masses of the order  $10 M_{\odot}$, at a semi-distance $R_{now} = L_0/2 = 5$ AU, the time for merging will be $t_{coal} \sim 10^9$ yr. This implies a (dimensionless) strain \footnote{See \cite{sensitivity} for details on different quantities for expressing  the sensitivity curve of GWs detectors. } of the order $10^{-50}$ for a source at a distance $r=100$ Mpc. A more promising scenario is offered by the merging of supermassive black holes, as is the case of SDSS J$153636.22+044127.0$ at a redshift $z\simeq 0.4$ (i.e. $\simeq 1700$ Mpc) \citep{colpi_2009}. With masses of $10^9 M_{\odot}$ and $10^7 M_{\odot}$ and a semi-distance of $0.05$ pc, the frequency\footnote{The frequency $f$ can be estimated remembering that the velocity is $v\simeq \sqrt{G M_{tot} / L}$ and $f \simeq v / (\pi L) $. More precisely, the frequency of the gravitational radiation would be twice that value.} and coalescence time would be $\sim 10^{-9}$ Hz and  $\sim 4 \times 10^{8}$ yr  respectively, thus implying an amplitude of $10^{-34}$, which is higher than the stellar-mass  black-hole binaries, but still very faint for all current and future gravitational wave observatories, including advanced LIGO, SKA, LISA and the Einstein Telescope, regardless of the value of the non-local parameter.  The most prosperous case would be  that of two black holes of 100 $M_{\odot}$  at an initial distance of $L_0 = 0.01 $ AU \footnote{The ratio between the Schwarzschild radius $R_s$, corresponding to the total mass of the system,  and the orbital separation $L_0$ is an indicator of the gravitational strength. With our assumptions, it gives $R_s/L_0 \simeq 0.04$. For a self-gravitating system, the virial theorem implies $v/c \sim \sqrt{R_s/d} \simeq 0.2$, meaning that relativistic corrections to the Newtonian motion of the binary system, although not negligible, do not affect very much our estimates. At closer distances, i.e. very near to the merging, significant corrections or  numerical relativity techniques should be adopted.}.  The frequency, in this case, would be $\sim 2 $ Hz. From Fig. \ref{fig1}, it is clear that  the dimensionless strain explodes as soon as $a \rightarrow 1/6$,  where the theory presents  a divergence.  


In a very narrow range,  for the non-local coupling constant,  it  could fall within the low frequency  sensitivity of the Einstein Telescope, thus representing a case of scientific interest for the detection of a massless transverse scalar mode in the very near future. See Fig.\ref{fig1}.

\section{Discussion and Conclusions}
\label{sec4}
Although a theory of quantum gravity remains unknown, it is legitimate to ask what happens when a strong gravitational field is present at quantum scales. Because of its weakness compared to  other interactions,  the characteristic scale under which one would
expect to experience non-classical effects relevant to gravity, the Planck
scale, is quite small, i.e. $10^{-33}$ cm, which is  not  accessible by any current
experiment. It is not a new fact that Quantum Mechanics shows non-local aspects: non-locality is a manifestation of entanglement and the latter has been repeatedly demonstrated in laboratory experiments; Bell's theorem, furthermore, demonstrated that locality is violated in  quantum systems. In gravitational physics,  non-locality mainly  emerges as a 'side' effect of the introduction of quantum corrections to GR in view  to cure the singularity problem. Indeed, applications to Cosmology showed that non-local ghost-free higher-derivative
modifications of the Einstein gravity in the ultraviolet regime can  admit non-singular bouncing solutions for the Universe (in place of the Big Bang singular  solution) and non-singular Schwarzschild metrics for black holes \cite{Modesto_2011}. Another motivation for considering non-local gravity is the possibility to achieve renormalizability without the appearance of ghost modes \cite{Modesto_Renormalizab}.  
More precisely, non-local theories of gravity   can be considered as Extended Theories of Gravity \cite{Capozziello:2011et}, meaning  that when the non-local terms are canceled out,  GR is restored. \\
In this paper, we studied a possible implementation of  non-local  gravity consisting in the addition of a non-local term of the form $R \Box^{-1}R$ to the Hilbert-Einstein Lagrangian. We derived, from it,  the power emitted by a gravitational system in terms of the quadrupole radiation, in view of possible future astrophysical detections.  In particular,  we obtained  the field equations  (using a localized version of the Lagrangian) of the theory and  solved  them in terms of the (source) stress-energy tensor $T_{\mu\nu}$, which, in the far region condition, only depends on the quadrupole momenta. Finally, from the expression of the  Noether current, we  computed the gravitational wave stress-energy pseudo-tensor, (GW-SET), $\tau_{\mu\nu}$, finding that the   GR quadrupole formula is modified in a non-trivial way. This fact
make it feasible to find  possible signatures of non-locality. Incidentally, we have found no mass connected to non-local terms and this implies no length scale associated to non-locality, at least in the class of theories considered here. This is different from the other class  of non-local gravity theories, namely the so called Infinite Derivative Gravity theories, where it results a lower bound on the non-local energy scale of $M \sim 10^{14}$ GeV \cite{10_14}.   Even if we  expect a very small value for the non-local parameter $a$ (any correction to GR in any extended theory of gravity should be rather slight), a preliminary application to a simple astrophysical scenario (a Keplerian binary system in circular orbit) showed that  non-local corrections are compatible with GR predictions, but the deviation cloud be, at least in principle, appreciable, paving the way for new observational constraints on non-local corrections compared with those already present in the literature \cite{Amendola_2019,Bouche2022, S2, Belgacem:2019lwx}. \\
In a forthcoming paper, we will apply our final result  (\ref{p_tot_finale}) to realistic astrophysical scenarios  (such as the Hulse-Taylor binary system) in order to  find constraints on the non-local parameter $a$. Another point to focus on will be considering the power emitted by a binary system in an  \textit{elliptical}
orbit, with the aim of computing how the emitted power depends on the eccentricity of the orbit in  presence of non-local corrections. We expect a strong dependence on eccentricity, as in GR, but a different enhancement factor. Other possible continuations of this work could concern  the extension to higher order corrections like $R\Box^{-n}R$ with $n>1$ \cite{Capriolo}. In particular, models  with $R\Box^{-2}R$-corrections seems to be very promising as discussed in \cite{Maggiore_DE_NonLocal,Maggiore_Redshift_2019}. 

\appendix

\section{Appendix}
The Lagrangian density for our model of non-local gravity in Eq. (\ref{action2}) is similar to a non-minimally coupled scalar-tensor theory. The presence of  scalar fields allows us to avoid integro-differential equations, relatively simplifying the calculations. However, the latter are very involved and long; so in this appendix, we report only the main steps that led first to Eq. (\ref{L2_final}) and then to Eq. (\ref{eq:2.19}). \\
The starting point is to write the second-order Lagrangian density, as given in Eq. (\ref{eq:sec}). It is

\begin{equation}
\label{A1}
    \mathcal{L}^{(2)} = \dfrac{1}{2 \chi} \Big[ (\sqrt{-g}R)^{(2)} \Big(1+a \phi_{0} -\lambda_{0}\Big) + R^{(1)} \Big(a\varphi -w\Big) - \partial^{\mu}\varphi \partial_{\mu}w \Big] \; ,
\end{equation}
where  $\psi_0 = 1+a\varphi -\lambda_0$ and $\psi^{(1)} = a\varphi - w$.  The first term in (\ref{A1}) is the second-order correction of the Ricci scalar. We have
\begin{equation}\label{A2}
 \dfrac{(\sqrt{-g}R)^{(2)}}{2\chi} = \dfrac{1}{64 \pi G} \Big[ \partial_{\alpha} h \partial^{\alpha}h + 2 \partial_{\alpha}h_{\beta \gamma} \partial^{\beta}h^{\alpha \gamma} -2 \partial^{\alpha}h\partial_{\beta}h^{\beta}_{\alpha}-\partial_{\gamma}h_{\alpha \beta}\partial^{\gamma}h^{\alpha \beta} \Big]\, . 
\end{equation}
It has to be rewritten  in terms  of the  gauge field $\theta_{\alpha \beta}$  given in Eq. (\ref{gauge}). The second term of Eq. (\ref{A1}), written  in  $\theta_{\alpha \beta}$,  is

\[\arraycolsep=1.4pt\def\arraystretch{2.2}
\begin{array}{cc}
 \dfrac{R^{(1)}\psi^{(1)} }{2 \chi}  = -\partial_{\nu}h^{\mu\nu}\partial_{\mu}\psi^{(1)}+\partial_{\alpha}h \partial^{\alpha}\psi^{(1)}  =    \\
 -\dfrac{1}{32 \pi G} \Big[ 2 \partial_{\alpha}\psi^{(1)} \partial_{\beta}\theta^{\alpha\beta} + \partial_{\mu}\psi^{(1)} \partial^{\mu}\theta + \dfrac{6}{\psi_0} \partial_{\mu}\psi^{(1)} \partial^{\mu}\psi^{(1)}    \Big]\ ,
\end{array}
\]
where in the first line we  integrated by parts, while, in the second line, we used the definition of $\theta_{\mu\nu}$. Putting all together in Eq. (\ref{A1}), we have

\begin{equation}
\dfrac{32 \pi G}{\psi_0} \mathcal{L}^{(2)} = \dfrac{1}{4}\partial_{\alpha}\theta \partial^{\alpha}\theta  + \partial_{\alpha}\theta_{\beta \gamma} \partial^{\beta}\theta^{\alpha \gamma} - \dfrac{3}{\psi_0^2}\partial_{\alpha}\psi^{(1)}\partial^{\alpha}\psi^{(1)} - \dfrac{1}{2} \partial_{\gamma}\theta_{\alpha\beta}\partial^{\gamma}\theta^{\alpha\beta} - \dfrac{2}{\psi_0}\partial_{\beta} \varphi \partial^{\beta}w \; .
\end{equation}

From the definition of $\psi^{(1)}$, it is  

\begin{equation}
  \partial_{\alpha}\psi^{(1)}\partial^{\alpha}\psi^{(1)} =   a^2 \partial_{\alpha}\varphi \partial^{\alpha}\varphi + \partial_{\alpha}w \partial^{\alpha}w - 2a \partial_{\alpha}w\partial^{\alpha}\varphi\ ,
\end{equation}

and hence the final expression for the second order Lagrangian $\mathcal{L}^{(2)}= \mathcal{L}^{(2)}(\theta_{\alpha\beta},\varphi,w)$ is given by

\[\arraycolsep=1.4pt\def\arraystretch{2.2}
\begin{array}{cc}
 \dfrac{32 \pi G}{\psi_0} \mathcal{L}^{(2)} =    \dfrac{1}{4}\partial_{\alpha}\theta \partial^{\alpha}\theta  + \partial_{\alpha}\theta_{\beta \gamma} \partial^{\beta}\theta^{\alpha \gamma} - \dfrac{1}{2} \partial_{\gamma}\theta_{\alpha\beta}\partial^{\gamma}\theta^{\alpha\beta} +  \\
 \dfrac{(6a-2\psi_0)}{\psi_0^2}\partial_{\alpha}w\partial^{\alpha}w - \dfrac{3}{\psi_0^2}\Big( a^2 \partial_{\alpha}\varphi\partial^{\alpha}\varphi  + \partial_{\alpha}w\partial^{\alpha}w \Big) \; ,
\end{array}
\]
which is equal to Eq. (\ref{L2_final}). Now, we notice that $\frac{\partial \mathcal{L}^{(2)}}{\partial \theta_{\alpha \beta}} = 0$, while

\begin{equation}
\dfrac{\partial \mathcal{L}^{(2)}}{\partial(\partial_{\gamma} \theta_{\alpha\beta})} = 2 \partial^{( \beta }\theta^{\alpha ) \gamma} - \partial^{\gamma}\theta^{\alpha\beta} + \dfrac{1}{2}\eta^{\alpha\beta} \partial^{\gamma}\theta \; ,
\end{equation}
and hence the Euler-Lagrange equation for the field $\theta_{\alpha\beta}$ is 

\[\arraycolsep=1.4pt\def\arraystretch{2.2}
\begin{array}{cc}
 \dfrac{\partial \mathcal{L}^{(2)}}{\partial \theta_{\alpha \beta}} - \partial_{\gamma}\dfrac{\partial \mathcal{L}^{(2)}}{\partial(\partial_{\gamma} \theta_{\alpha\beta})} = 0 \; , \\ 
 - \Box \theta^{\alpha\beta} + \dfrac{1}{2}\eta^{\alpha\beta} \Box \theta = 0  \; \; \; \; \; \; \; \; \; \Rightarrow \; \; \; \; \; \; \;  \; 
 \Box \theta^{\alpha \beta} = 0 \; . 
\end{array}
\]

where in the second line we  took advantage of the Lorentz gauge and  exploited the trace equation. In a similar way, we arrive at the other two equations of motion (\ref{eq:eomphi}) and (\ref{eq:eomw}) for the remaining fields $\varphi$ and $w$. In particular, for this purpose, we needed the following partial derivatives: 

\[\arraycolsep=1.4pt\def\arraystretch{2.5}
\begin{array}{cc}
    \dfrac{\partial \mathcal{L}^{(2)}}{\partial(\partial_{\gamma}\varphi)} = \dfrac{\partial}{\partial(\partial_{\gamma}\varphi)} \Big[ \dfrac{(6a-2\psi_0)}{\psi_0^2} \partial_{\alpha}\varphi \partial^{\alpha}w - \dfrac{3}{\psi_0^2} a^2  \partial_{\alpha}\varphi \partial^{\alpha}\varphi \Big] = \\
    \dfrac{(6a-2\psi_0)}{\psi_0^2}\delta^{\gamma}_{\alpha}\partial^{\alpha}w - \dfrac{6}{\psi_0^2}a^2\partial^{\gamma}\varphi\ ,
\end{array}
\]

and 

\[\arraycolsep=1.4pt\def\arraystretch{2.5}
\begin{array}{cc}
    \dfrac{\partial \mathcal{L}^{(2)}}{\partial(\partial_{\gamma}w)} = \dfrac{\partial}{\partial(\partial_{\gamma}w)} \Big[ \dfrac{(6a-2\psi_0)}{\psi_0^2} \partial_{\alpha}\varphi \partial^{\alpha}w - \dfrac{3}{\psi_0^2} a^2  \partial_{\alpha}w \partial^{\alpha}w\Big] = \\
    \dfrac{(6a-2\psi_0)}{\psi_0^2}\delta^{\gamma}_{\alpha}\partial^{\alpha}\varphi - \dfrac{6}{\psi_0^2}\partial^{\gamma}w \; .
\end{array}
\]

The above expressions  have also been used in Eq. (\ref{curr}), which becomes 

\[\arraycolsep=1.4pt\def\arraystretch{2.5}
\begin{array}{cc}
    j^{\alpha}_{\beta}= \dfrac{\psi_0}{32 \pi \varphi}\Big(-2 \partial^{\nu} \theta^{\mu \alpha}+\partial^{\alpha} \theta^{\mu \nu }\Big)  \partial_\beta \theta_{\mu v} -\dfrac{\psi_0}{32 \pi G}\Big[\frac{(6 a_1-2 \psi_0)}{\psi_0^2} \partial^\alpha w 
    -\dfrac{6}{\psi_0^2} a^2  \partial^\alpha \varphi \Big] \partial_\beta \varphi \\ 
    -\dfrac{\psi_0}{32 \pi G}\Big[\dfrac{(6a-2 \psi_0)}{\psi_0^2}  \partial^\alpha \varphi-\dfrac{6}{\psi_0^2} \partial^\alpha w \Big] \partial_\beta w  
   +  \dfrac{\psi_0\delta_\beta^\alpha}{32 \pi G}\Big[\dfrac{1}{4}\partial_\alpha\theta\partial^{\alpha}\theta +    \partial_\nu \theta_{\mu \gamma}  \partial^\gamma \theta^{\mu \nu}-\dfrac{1}{2} \partial_\gamma \theta_{\mu \nu}  \partial^\gamma \theta^{\mu \nu} \\
   +\dfrac{(6 a-2 \psi_0)}{\psi_0^2} \partial^\gamma \varphi  \partial_\gamma w 
   -\dfrac{3}{\psi_0^2}\Big(a^2  \partial_\mu \varphi  \partial^\mu \varphi+\partial_\mu w \partial^\mu w \Big)\Big] \; .
\end{array}
\]

Using the Lorenz gauge condition, the field equations and integrating by parts when necessary,  the above equations  simplify to

\[\arraycolsep=1.4pt\def\arraystretch{2.2}
\begin{array}{cc}
   j^{\alpha}_{\beta} = \dfrac{\psi_{0}}{32 \pi G} \Big[     \partial^{\alpha}\theta^{\mu\nu}\partial_{\beta}\theta_{\mu\nu} - K\partial^{\alpha} w \partial_{\beta}\varphi 
   
   + \dfrac{6}{\psi_{0}^2} a^2 \partial^{\alpha}\varphi \partial_{\beta}\varphi - K\partial^{\alpha} \varphi \partial_{\beta}w 
   + \dfrac{6}{\psi_0^2} \partial^{\alpha} w \partial_{\beta}w \\
  + \delta^{\alpha}_{\beta} K \partial^{\gamma} \varphi \partial_{\gamma}w - \delta^{\alpha}_{\beta} \dfrac{3 a^2}{\psi_0^2} \partial^{\gamma} \varphi \partial_{\gamma}\varphi - \delta^{\alpha}_{\beta}  \dfrac{3 }{\psi_0^2} \partial^{\gamma} w \partial_{\gamma}w  \Big] \ ,
\end{array}
\]
where  we defined $K = (6a-2\psi_{0})/\psi_{0}^{2}$. After noticing that the last three terms are zero as well (see Sec. \ref{sec2}), we easily  obtain Eq. (\ref{eq:2.19}).

\section{Appendix}
In this appendix, we briefly report the calculations  leading  to  Eq.  (\ref{power_final_NL}). 
We need  to write Eq. (\ref{pseudo_final_new}), i.e.

\begin{equation}\label{pseudo_final_new2}
    \tau^{\alpha}_{\beta} = \dfrac{\psi_0}{32 \pi G } \Big\langle   \partial^{\alpha} \theta^{\mu\nu}_{(TT)}  \partial_{\beta} \theta_{\mu\nu}^{(TT)} + 
    4a(6a-1) \partial^{\alpha}\varphi \partial_{\beta} \varphi \Big\rangle  \; ,
\end{equation}

in terms of $J_{\mu\nu}$, defined in Eq. (\ref{eq:38}), and then $J_{\mu\nu}$ in terms of the quadrupole momentum $Q_{ij}$, defined in Eq. (\ref{quadrupoleQ}). First, we expand $J_{\mu\nu}$ in Taylor series around $t'=t-r$, i.e. \cite{Laurentis_2011}

\begin{multline}\label{J_Taylor}
    J^{\mu \nu}(\mathbf{x}, t)=\frac{4}{r}\Big[\int d^3 \mathbf{x}^{\prime} T^{\mu \nu}\left(\mathbf{x}^{\prime}, t^{\prime}\right)
+\hat{x} \int d^3 \mathbf{x} \mathbf{x}^{\prime} \frac{\partial T^{\mu \nu}(\mathbf{x}, t)}{\partial t^{\prime}} \\
+\frac{1}{2} \int d^3 \mathbf{x}^{\prime}(\hat{x} \cdot \mathbf{x}^{\prime})^2 \frac{\partial^2 T^{\mu \nu}(\mathbf{x}, t)^2}{\partial t^{\prime}}\Big],
\end{multline}

where we assumed that, in the far region, it is $|\mathbf{x}-\mathbf{x}'|^{-1} \simeq \frac{1}{r}$ and $|\mathbf{x}-\mathbf{x}'| \simeq r - \hat{x}\cdot \mathbf{x}'$, with $r=|\mathbf{x}|$.
Using Eq. (\ref{theta_quadr}) rewritten in the TT gauge, one finds that the GR contribution to the GW-SET is   
\begin{equation}
\tau_{0r}^{(GR)} = \dfrac{1}{32 \pi G} \langle \partial_{0}\theta^{ij}_{(TT)} \partial_{r}\theta_{ij}^{(TT)} \rangle = - \dfrac{G}{8 \pi r^2} \langle \dddot{Q}_{ij}^{(TT)} \dddot{Q}^{ij}_{(TT)}  \rangle    \; .
\end{equation}
Following the standard computations in GR \cite{Carroll:2004st,Maggiore:I}, one immediately\footnote{One needs to write $Q_{ij}^{(TT)}$ as $Q_{ij}^{(TT)}=\bar{Q}_{ij}^{(TT)}$ where $\bar{Q}_{ij}=Q_{ij}-\frac{1}{3}\delta_{ij}\delta^{kl}Q_{kl}$ and then use the projector tensors (see Chapter 7 in \cite{Carroll:2004st}) to expand the TT gauge components.    } obtains Eq. (\ref{powerGR}). \\

For the second term of Eq. (\ref{pseudo_final_new2}),  we use Eq. (\ref{phi_integrale}) and  notice that, in our metric signature, it is  $J=J^{\mu}_{\mu} = J^{00}-J^{11}-J^{22}-J^{33} $.  Using Eq. (\ref{J_Taylor}), after long but straightforward computations,   we get Eq. (\ref{J}), whose time derivative reads as    

\begin{equation}
    \dot{J}(t') = \dfrac{2G}{r} \Big[  n_i n_j  \dddot{Q}^{ij} + \dddot{Q}  \Big]\ ,
\end{equation}

where we used mass and angular momentum conservation. It must be emphasized that although usually the component $J_{00}$ is approximated as $J_{00} \simeq \frac{4GM}{r} + \mathcal{O}\big(\frac{1}{r^3} \big)$, here it is necessary to  retain  higher order terms as well  in order not to lose its  quadrupole contribution,

\begin{equation}
    J_{00} \simeq \dfrac{4G}{r} \big[M + n_i \dot{D}^i + \dfrac{1}{2} n_i n_j \ddot{Q}^{ij} \Big] \;.
\end{equation}

Since for the computation of the gravitational power $P_{tot}$ only the component $\tau_{0r}$ is involved, we just need to compute the term $\partial_0 \varphi \partial_r \varphi$, with $\varphi$ given by Eq. (\ref{phi_integrale}). In particular, it will be

\begin{equation}
    \partial_0 J= \dfrac{4G}{r} \Big[ \dfrac{1}{2} n_i n_j \dddot{Q}^{ij} \Big ] + \dfrac{2G}{r} \dddot{Q}\ ,
\end{equation}
and
\begin{equation}
  \partial_r J \simeq - \dfrac{2G}{r}n_i n_j   \dddot{Q}^{ij} - \dfrac{2G}{r} \dddot{Q}\ ,
\end{equation}

where in  $\partial_r J$ we neglected $1/r^2$-terms. Putting all together we obtain Eq. (\ref{part_part}). Therefore,  to arrive at   Eq. (\ref{power_final_NL}), we need to perform the integral

\begin{equation}
 P_{tot}^{(NL)} = \dfrac{a G}{8 \pi (6a-1)} \int d\Omega \Big[ n_i n_j \dddot{Q}_{ij} + \dddot{Q}  \Big]^2 \ ,  
\end{equation}
which explicitly reads as

\begin{equation}
P_{tot}^{(NL)} = \dfrac{a G}{8 \pi (6a-1)} \Big[  \dfrac{4\pi}{15}  \Big(\delta_{ij}\delta_{kl}+\delta_{ik}\delta_{jl}+\delta_{il}\delta_{jk} \Big) \dddot{Q}^{ij} \dddot{Q}^{kl} + 4\pi\dddot{Q}^2 + \dfrac{8\pi}{3} \delta_{ij}\dddot{Q}^{ij}\dddot{Q}^2 \Big]\ ,
\end{equation}
and simplifying

 \begin{equation}\label{fine}
    P_{tot}^{(NL)} = \dfrac{a G}{8 (6a-1)} \Big\langle  \dfrac{8}{15} \dddot{Q}^{jk}\dddot{Q}_{jk} + \dfrac{104}{15} \dddot{Q}^2 \Big\rangle \; . 
\end{equation}

It is evident that, this equation, after a trivial simplification, reduces to  Eq. (\ref{power_final_NL}).

\acknowledgments

The authors acknowledge the Istituto Nazionale di Fisica Nucleare (INFN), sezione di Napoli, \textit{iniziative specifiche} QGSKY and MOONLIGHT-2, and the Istituto Nazionale di Alta Matematica (INdAM), Gruppo Nazionale di Fisica Matematica. 
This paper is based upon work from COST Action CA21136 {\it Addressing observational tensions in cosmology with systematics and fundamental physics} (CosmoVerse) supported by COST (European Cooperation in Science and Technology).




\bibliography{biblio2.bib}
\bibliographystyle{apsrev}


\end{document}